Hindawi

## Review Article

# Role of Magnetic Interaction in Dense Plasma

**S. Sarkar,[1] K. Pal,[2] and A. K. Dutt-Mazumder[1]**

[1] *High Energy Nuclear and Particle Physics Division, Saha Institute of Nuclear Physics, 1/AF Bidhannagar,*
  *Kolkata 700 064, India*
[2] *Serampore College, Serampore 712201, India*

Correspondence should be addressed to S. Sarkar; sreemoyee.sarkar@saha.ac.in

Quasiparticle excitations and associated phenomena of energy and momentum transfer rates have been calculated in terms of the drag and the diffusion coefficients exposing clearly the dominance of the magnetic interaction over its electric counterpart. The results have been compared with the finite temperature results highlighting the similarities and dissimilarities in the two extreme regimes of temperature and density. Non-Fermi-liquid behavior of various physical quantities like neutrino mean free path and thermal relaxation time due to the inclusion of magnetic interaction has clearly been revealed. All the results presented in the current review are pertinent to the degenerate and ultradegenerate plasma.

## 1. Introduction

Understanding the properties of the hot and dense ultrarelativistic plasma has been at the forefront of contemporary research for the past few decades. Interests in this regime cover a broad area starting from the laboratory based heavy ion collisions (HIC) to the wider domain of naturally occurring astrophysical sites like neutron stars, supernovae, and white dwarfs and so forth.

In the present review we mainly focus on the properties of plasma with high chemical potential ($\mu$) and zero or low temperature ($T$). It is to be noted that the theoretical calculations so far involving both the quantum electrodynamics and quantum chromodynamics (QED/QCD) have been confined largely to the domain of high temperature particularly to address the issues related to heavy ion collisions. The main motivation for this has been the possibility of quark-hadron phase transition which might occur in HIC mimicking the conditions of microsecond old universe. Studies with dense plasma on the other hand find their applications mainly in astrophysics.

Several calculations have recently been performed exposing subtle departure of the characteristic behavior of the quasiparticle excitations in dense plasma than their high temperature counterparts. For example, the role of magnetic interaction in dense plasma, as we shall see, modifies the fermion self-energy in a nontrivial way involving fractional powers of the excitation energy [1] leading to phenomena not seen in the high temperature plasma. In fact the transverse or magnetic interaction seems to dominate over the corresponding electric or longitudinal interactions [1] in degenerate electron or quark matter. It is to be noted that such departures can only be significant in the relativistic plasma as in the nonrelativistic case magnetic interactions are $\beta = v/c$ suppressed [2–5]. The introduction of magnetic interaction changes some of the characteristic behaviours of dense plasma which show departure from the normal Fermi liquid case. In this review this particular aspect will be illuminated upon further.

Our discussion starts with the quasiparticle damping rate ($\gamma$). In the relativistic case $\gamma \sim (E - \mu)$ [6] unlike the known non-relativistic scenario where $\gamma \sim (E - \mu)^2$. Here, $E$ is the energy of the quasiparticle and $\mu$ is the chemical potential [6–8]. More importantly the infrared behavior of this quantity differs dramatically from the high temperature case [6–8]. In a plasma the other quantities of interest have been the drag ($\eta$) and the diffusion coefficients ($\mathcal{B}$) which eventually determine the equilibration time scale of the plasma. These two quantities are actually related to the energy loss and the momentum transfer due to the scattering of the constituent particles of the plasma. In high temperature case they are known to be connected *via* the Einstein relation. We address this issue to see the scaling behavior of these quantities in the domain of zero and high temperature plasma [2].

Recently the implications of the medium modified quasi-particle self-energies have been discussed by several authors showing their significance in the cooling behavior of neutron stars. The appearance of logarithmic terms in the specific heat, emissivity, and thermal relaxation time of the degenerate matter has been found to be of crucial importance. How the medium modified dispersion relation changes the cooling behavior of neutron star has been first exposed in [9] and later in [3, 5, 10].

The plan of the review is as follows. In Section 2 we have shown the behavior of quasiparticle damping rate in ultradegenerate plasma. In this section we compare the results of quasiparticle lifetime in the extreme limits of high temperature and high density. For the former it is known that the damping rate for the exchange of static gauge boson remains unscreened even when one uses hard thermal loop corrected propagator for the gauge bosons [11, 12]. It has been shown in the literatures that a further resummation is required to resolve this issue by invoking Bloch-Nordseick techniques as discussed in great detail in [13, 14]. For the degenerate case, as we shall see this problem does not appear. In this context however we shall expose the relative importance of the magnetic interaction over the electric part unlike the hot relativistic plasma.

In this review we also calculate the drag and the diffusion coefficients for both degenerate and ultradegenerate cases. Different properties of $\eta$ and $\mathcal{B}$ in the zero temperature and low temperature limit have been reported in Section 3. In Section 4 explicit analysis has been presented to exhibit how the nature of neutrino mean free path changes with the inclusion of the transverse mode. One distinguishing feature of the present work is to show the structure of the fermion self-energy near the Fermi surface. Of particular importance is the vanishing group velocity at the Fermi surface which in turn is responsible for the non-Fermi liquid (NFL) corrections. Section 5 describes the implication of the neutrino mean free path *vis-a-vis* cooling of the neutron star specially focusing on NFL corrections to the various physical quantities. We calculate the thermal relaxation time exposing the role of magnetic interaction further in Section 6 where we also discuss NFL corrections.

In the last two sections of this review we estimate the neutrino mean free path and thermal relaxation time of degenerate matter. At very high density nowadays it is well known that quark matter is expected to form a color superconducting color-flavour locked (CFL) phase [15]. In this phase all quark excitations are gapped, and the mean free path or the thermal relaxation time is modified by exponential factor involving gap parameter [16–18]. However, in order to highlight particular aspects of NFL behaviour of the relativistic plasma we deal with only ungapped quark

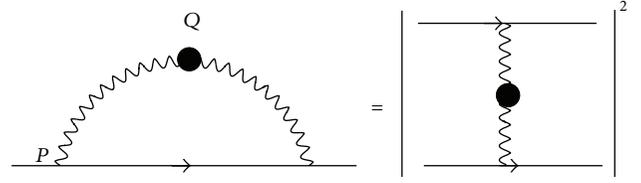

Figure 1: Fermion self-energy with resummed photon propagator.

matter. The extension of the present calculation for the gapped quark matter is straightforward.

## 2. Damping Rate

The propagation of a particle immersed in plasma gets modified through interaction with the medium. In case of ultradegenerate ($T = 0$, $\mu \neq 0$) plasma these medium modified particles (quasiparticles) scatter with the other particles close to the Fermi surface. Information about the quasiparticle lifetime is obtained from the retarded propagator. The damping rate is expected to follow exponential decay in time $S_R(t, p) \sim e^{-Et}e^{-\gamma t}$, so that $|S_R(t, p)|^2 \sim e^{-\Gamma t}$, where $\Gamma = \gamma/2$. The life time of the single-particle excitation is then $\tau(p) = 1/\Gamma(p)$. The concept of quasiparticles becomes meaningful only if their life time is long or in other words the damping width is much smaller than its energy. In a plasma the damping rate of the quasiparticle can be written in terms of the imaginary part of the fermion self-energy or the scattering rate (see Figure 1). Mathematically one writes [19]

$$\gamma(E) = -\frac{1}{4E} \text{Tr} \left[ \text{Im} \, \Sigma \left( p_0 + i\sigma \mathbf{p} \right) \not{P} \right] \Big|_{p_0 = E}, \quad (1)$$

where $\Sigma$ is the fermion self-energy, $\sigma \rightarrow 0^+$, and we have used $P_\mu = (p_0, \mathbf{p})$. The aforementioned expression is true for both the high temperature and the zero temperature plasmas. The one-loop electron self-energy is given by

$$\Sigma(P) = e^2 \int \frac{d^4q}{(2\pi)^4} \gamma_\mu \, S_f(P - Q) \, \gamma_\nu \, \Delta_{\mu\nu}(Q). \quad (2)$$

In the previous equation $S_f(P - Q)$ is the free fermion propagator and $\Delta_{\mu\nu}(Q)$ is the photon propagator.

Focusing on (2) we can see that it involves the boson propagator ($\Delta_{\mu\nu}$) which is well known to diverge in the infrared regime. This is a generic feature of both the high temperature and the zero temperature plasmas. In case of hot QED or QCD plasmas one handles this problem by employing the hard thermal loop (HTL) prescription which was originally developed in [11, 12]. In this method the exchanged momentum (**q**) integration is performed in two domains by introducing an intermediate cut-off parameter ($q^*$), below which the one-loop corrected dressed boson propagator has to be used and the bare propagator can be used to perform the integration above $q^*$ (Braaten and Yuan's prescription) [20].

For electric interaction this is sufficient to remove the infrared singularity associated with the exchange of massless bosons (photons or gluons). The interaction mediated by the transverse bosons on the other hand still poses a problem at high temperature [13, 14]. We shall discuss this further at the end of this section where we compare the zero temperature case with the corresponding high temperature calculations. It might be mentioned that to construct the one-loop corrected photon propagator one needs to construct the photon self-energy in dense plasma [21]. In the ultradegenerate case, the dominant contribution to the photon self-energy comes from the loop momentum $\sim\mu$, and the external momentum for the soft modes is assumed to be $\mathbf{q} \sim e\mu$ (Figure 2). This, in principle, runs similar to the HTL approximation where the loop momentum is $\sim T$ and the external momentum is $\sim eT$. Likewise for the degenerate case one can construct hard dense loop (HDL) propagator where loop momentum is assumed to be $\sim\mu$ [21].

The structure of the dressed photon propagator in the Coulomb gauge can be written as [19]

$$\Delta_{\mu\nu}(Q) = \delta_{\mu 0}\delta_{\nu 0} \; \Delta_l(Q) + P^t_{\mu\nu}\Delta_t(Q), \qquad (3)$$

with $P^t_{ij} = (\delta_{ij} - \hat{q}_i\hat{q}_j)$, $\hat{q}^i = \mathbf{q}^i/|\mathbf{q}|$, $P^t_{i0} = P^t_{0i} = P^t_{00} = 0$, and $\Delta_l$, $\Delta_t$ are given by [19]

$$\Delta_l(q_0, q) = \frac{-1}{q^2 + \Pi_l}, \qquad (4)$$

$$\Delta_t(q_0, q) = \frac{-1}{q_0^2 - q^2 - \Pi_t}. \qquad (5)$$

$\Delta_{l,t}$ are the longitudinal and transverse components of the boson propagator, and $\Pi_{l,t}$ are the longitudinal and transverse parts of the self-energy [19]:

$$\Pi_l(q_0, q) = m_D^2 \left[ 1 - \frac{q_0}{2q} \ln\left(\frac{q_0 + q}{q_0 - q}\right) \right],$$

$$\Pi_t(q_0, q) = m_D^2 \left[ \frac{q_0^2}{2q^2} + \frac{q_0\left(1 - (q_0^2/q^2)\right)}{4q} \ln\left(\frac{q_0 + q}{q_0 - q}\right) \right], \qquad (6)$$

where the Debye mass is $m_D = e\mu/\pi$.

The bosonic spectral functions are the imaginary part of the propagators:

$$\rho_{l,t}(q_0, \mathbf{q}) = 2\operatorname{Im}\Delta_{l,t}(q_0 + i\sigma, \mathbf{q}), \qquad (7)$$

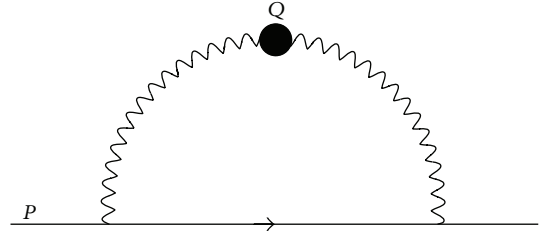

Figure 2: Feynman diagram for $e^-e^- \rightarrow e^-e^-$ scattering process with effective photon propagator.

where $\rho_{l,t}$ in the HDL approximations are given by [6]

$$\rho_l(q_0, q)$$
$$= \frac{2\pi m_D^2 x\Theta\left(1 - x^2\right)}{2\left[q^2 + m_D^2\left(1 - (x/2)\ln\left|(x+1)/(x-1)\right|\right)\right]^2 + \left(m_D^4\pi^2 x^2/4\right)},$$

$$\rho_t(q_0, q)$$
$$= 2\pi m_D^2 x\left(1 - x^2\right)\Theta\left(1 - x^2\right)$$
$$\times \left(\left[2q^2\left(x^2 - 1\right) - m_D^2 x^2\left(1 + \frac{(1 - x^2)}{2x}\ln\left|\frac{x+1}{x-1}\right|\right)\right]^2 \right.$$
$$\left. + \frac{m_D^4\pi^2 x^2\left(1 - x^2\right)^2}{4}\right)^{-1}, \qquad (8)$$

with $x = q_0/q$.

To calculate $\gamma$ in the ultradegenerate plasma the energy of the quasiparticle is therefore considered to be hard, that is, $E \sim \mu$. Hence, for the quasiparticle with momentum close to the Fermi momentum the electron-photon vertex can be replaced by the bare one and the one-loop self-energy is dominated by the photon with soft momentum ($\mathbf{q} \sim e\mu$). So, the calculation of imaginary part of $\Sigma$ can be performed with the help of the free fermion spectral function $\rho_f(P - Q) = 2\pi\varepsilon(k_0)\delta(k_0^2 - E_k^2)$ ($\varepsilon(k_0) = k_0/|k_0|$) and the medium modified $\rho_{l,t}$ [6]. Explicitly,

$$\gamma(E) = \frac{\pi e^2}{E} \int \frac{d^3q}{(2\pi)^3} \int_{-\infty}^{\infty} \frac{dk_0}{2\pi}\rho_f(k_0) \int_{-\infty}^{\infty} \frac{dq_0}{2\pi}$$
$$\times \left(1 + n(q_0) - \bar{n}(k_0)\right)\delta\left(E - k_0 - q_0\right)$$
$$\times \left\{\left[p_0 k_0 + \mathbf{p}\cdot\mathbf{k}\right]\right.$$
$$\left. \times \rho_l(q_0, q) + 2\left[p_0 k_0 - (\mathbf{p}\cdot\hat{\mathbf{q}})(\mathbf{k}\cdot\hat{\mathbf{q}})\right]\rho_t(q_0, q)\right\}. \qquad (9)$$

In the previous equation $n(q_0)$ and $\bar{n}(k_0)$ are the boson and fermion distribution functions. In case of the cold plasma these distribution functions can be replaced by $(1 + n(q_0)) = \Theta(q_0)$ and $\bar{n}(k_0) = \Theta(\mu - E + q_0)$, where $\Theta$ represents the step function. These theta functions, as we shall see, restrict the phase space of the $q_0$ integration severely. This eventually with the help of the HDL propagator makes $\gamma$ finite [13, 14, 21]. To proceed further an approximation in the region $(E-\mu)/m_D = \nu \ll 1$ has been made. This is due to the fact that scatterings close to the Fermi surface are only relevant as mentioned earlier. To perform the $q$ integration the integration domain is divided into two sectors. In the soft sector ($q < q^*$) with the HDL corrected spectral function mentioned in (8) one obtains [6]

$$\gamma_{\text{soft}}(E) \simeq \frac{e^2 m_D \nu}{24\pi} + \frac{e^2 m_D^4 \nu^2}{32\pi}\left(\frac{\pi}{2m_D^3} - \frac{1}{q^{*3}}\right). \qquad (10)$$

Only the leading order terms in $(E-\mu)/m_D$ have been retained in writing the previous equation. For the hard momentum region $q > q^*$ bare propagator is sufficient to calculate the damping rate:

$$\gamma_{\text{hard}}(E) \simeq \frac{e^2 m_D^4 \nu^2}{32\pi}\left(\frac{1}{q^{*3}} - \frac{1}{q_{\max}^3}\right), \qquad (11)$$

where $q_{\max} \simeq \mu$ is the maximum momentum transfer that is allowed by kinematics.

Two points are to be noted here. One that $q^*$ dependent term appear only at the higher orders that is, the leading order result is independent of $q^*$. Second upon summation the $q^*$ terms cancel out exactly. The total damping rate is then given by [6]

$$\gamma(E) \simeq \mathcal{G}_1 \nu + \mathcal{G}_2 \nu^2 + \cdots, \qquad (12)$$

where $\mathcal{G}_1 = (e^2 m_D)/(24\pi)$ and $\mathcal{G}_2 = (e^2 m_D)/32$. In the last equation the first term corresponds to the magnetic interaction, and the second term comes from the electric interaction revealing the fact that the magnetic interaction here dominates over the electric one. In this connection we can recall the results of the damping rate in case of the high temperature plasma with vanishing chemical potential. These are given by [13, 14]

$$\gamma_t = \frac{e^2 T}{2\pi}\int_0^{\omega_p}\frac{dq}{q}, \qquad \gamma_l \sim e^2 T, \qquad (13)$$

where $\omega_p$ is the plasma frequency related to the Debye mass ($\omega_p = m_D/\sqrt{3}$).

The last equations have been obtained by using the one-loop corrected dressed propagator which makes $\gamma_l$ finite. The transverse part, on the other hand, remains divergent that can be cured by another resummation as discussed in [13, 14]. In the case of dense QED or QCD matter interestingly the second resummation is not required. This happens, as indicated above, due to the restrictions imposed by the Pauli blocking which severely cuts off the phase space. A comparative study between high and zero temperature damping rates has been listed in Table 1 [22].

In the nonrelativistic case the excitation energy dependence of $\gamma$ can be derived from the simple phase-space factor without performing any rigorous calculation. A particle with energy $E > \mu$ interacts with the particle of energy $E_k < \mu$.

The scattered particles are now in the energy state $E_{p'} > \mu$ and $E_{k'} > \mu$. The total probability of the scattering process is proportional to $\int \delta(E + E_k - E_{p'} - E_{k'})d^3k\,d^3p'$. For $E - \mu \ll \mu$, the permissible regions of variation for the scattered particles are $\mu < E_{p'} < E + E_k - \mu$ and $2\mu - E < E_k < \mu$. The integral over $d^3k$ and $d^3p'$ can be computed with the approximation $E_k \approx E_{p'} \approx \mu$, that is, scattering occurs near the Fermi surface. The scattering probability is then $(E-\mu)^2$ [23]. Similar arguments can be made for the transverse sector which will render the damping rate of the particle $\sim (E-\mu)$ at the leading order for relativistic case.

## 3. Drag and Diffusion Coefficients

### 3.1. Zero Temperature Plasma.
Next we review the drag and diffusion coefficients in zero temperature relativistic plasma and see how the magnetic interaction changes the behavior of these coefficients. These two are the important quantities to study the equilibrium properties of the plasma as mentioned earlier and are related to the energy and the momentum transfer per scattering, respectively. $\eta$ can thus be defined as follows [24]:

$$\eta = \frac{1}{E}\left(-\frac{dE}{dx}\right). \qquad (14)$$

The energy loss $-(dE/dx)$ can be calculated by averaging over the interaction rate ($\Gamma$) times the energy transfer per scattering $\omega$ [25]:

$$\frac{dE}{dx} = \int d\Gamma\,\omega, \qquad (15)$$

where $\Gamma = \gamma/2$. Inserting the energy exchange $\omega$ in the expression of $\gamma$ to calculate $dE/dx$ from the previous equation we obtain (Figure 2)

$$\begin{aligned}
-\frac{dE}{dx} = \frac{\pi e^2}{E}&\int\frac{d^3q}{(2\pi)^3}\int_{-\infty}^{\infty}\frac{dk_0}{2\pi}\rho_f(k_0)\int_{-\infty}^{\infty}\frac{dq_0}{2\pi}q_0 \\
&\times (1 + n(q_0) - \bar{n}(k_0))\,\delta(E - k_0 - q_0) \\
&\times [p_0 k_0 + \mathbf{p}\cdot\mathbf{k}]\,\rho_l(q_0, q) \\
&+ 2[p_0 k_0 - (\mathbf{p}\cdot\hat{\mathbf{q}})(\mathbf{k}\cdot\hat{\mathbf{q}})]\,\rho_t(q_0, q).
\end{aligned} \qquad (16)$$

The basic assumption behind the calculation of $\eta$ lies in the fact that $(E-\mu)/m_D = \nu \ll 1$. Taking into consideration the previous assumption in case of the hard photon exchange we get [2],

$$\left(-\frac{dE}{dx}\right) \simeq \frac{e^2 \nu^3 m_D^5}{16\pi}\int\frac{dq}{q^4}. \qquad (17)$$

The above expression is infrared divergent with algebraic dependence. Here, it would be worthwhile to comment that in case of the high temperature plasma the divergence is logarithmic ($e^4 T^2/24\pi E)\int(dq/q)$. The divergence is even worse in case of the ultradegenerate plasma. But we show below that with the HDL propagator and Pauli blocking one

TABLE 1: Comparison of high and zero temperature damping rates.

|  | $\gamma$ in high temperature plasma | $\gamma$ in zero temperature plasma |
| --- | --- | --- |
| Fermion | $f(E) = \dfrac{1}{e^{\beta(E-\mu)} + 1}$ | $f(E) = \Theta(\mu - E + q_0)$ |
| Boson | $n(E) = \dfrac{1}{e^{\beta E} - 1}$ | $1 + n(E) = \Theta(q_0)$ |
| Longitudinal contribution | $\gamma_l \sim e^2 T$ | $\gamma_l \simeq \dfrac{e^2 v^2 m_D}{64}$ |
| Transverse contribution | $\gamma_t \sim \dfrac{e^2 T}{2\pi} \int_o^{q^*} \dfrac{dq}{q}$ | $\gamma_t \simeq \dfrac{e^2 v m_D}{24\pi}$ |
| Infrared behaviour | IR divergent after the HTL resummation | Finite with HDL resummation |
| Dominance | $\gamma_l$ and $\gamma_t$ contribute at the same order ($O(e^2)$) | Dominant contribution from $\gamma_t$ |

can render $\eta$ finite. In the soft sector $\eta$ is given by the following expression:

$$\eta|_{\text{soft}}(E)$$
$$\simeq \frac{e^2}{8\pi^2 E}$$
$$\times \int_D dq\, q \int dq_0\, q_0 \left\{ \rho_l(q_0, q) + \left(1 - \frac{q_0^2}{q^2}\right) \rho_t(q_0, q) \right\}. \tag{18}$$

The integration domain ($D$) above is restricted by the $\Theta$ functions:

$$D : 0 \le q_0 \le E - \mu; \tag{19}$$
$$q \le q \le q^*.$$

After explicit calculation, the electric and the magnetic contributions to the expressions of $\eta$ take the following form [2]:

$$\eta|_{\text{soft}}^l \simeq \frac{e^2 m_D^2 v^3}{96 E} - \frac{e^2 m_D^5 v^3}{72\pi q^{*3} E},$$
$$\eta|_{\text{soft}}^t \simeq \frac{e^2 m_D^2 v^2}{48\pi E} - \frac{e^2 m_D^5 v^3}{144\pi q^{*3} E}. \tag{20}$$

In the last two equations the leading order terms are independent of the cut-off parameter as they appear only at $O(e^7)$. In the region of hard photon momentum exchange we obtain [2]

$$\eta|_{\text{hard}} \simeq \frac{e^2 m_D^5 v^3}{48\pi E} \left[ \frac{1}{q^{*3}} - \frac{1}{\mu^3} \right]. \tag{21}$$

On addition of the soft and the hard sectors it can be seen that the expression of the drag coefficient becomes [2],

$$\eta \simeq \mathscr{H}_1 v^3 + \mathscr{H}_2 v^3 + \cdots, \tag{22}$$

where $\mathscr{H}_1 = (e^2 m_D^2)/(48\pi E)$ and $\mathscr{H}_2 = (e^2 m_D^2)/(96 E)$. In the previous equation the first term corresponds to the magnetic interaction, and the second term has come from the longitudinal interaction. From (21) it is important to note that the hard sector involves even higher power. The entire contribution at the leading order then comes from the soft sector alone, and Braaten and Yuan's prescription is not required in case of the ultradegenerate plasma [2]. The final expression of $\eta$ reveals that with the magnetic interaction $\eta$ changes significantly from the Fermi liquid result. In the Fermi liquid theory, since the magnetic interaction is suppressed in comparison with the electric sector $\eta$ goes with $(E-\mu)^3$. Here, we can recall the result of the drag coefficient in case of high temperature plasma and compare the same with the previous result. The coefficient at high temperature looks like [25–29]

$$\eta \simeq \frac{e^4 T^2}{36\pi E} \left( -\frac{1}{2} + \frac{3}{2} \log \left| \frac{q_{\max}}{m_D} \right| + \frac{1}{2} \log \left| \frac{2}{\sqrt{\pi}} \right| \right). \tag{23}$$

From the previous result it is clear that there is no splitting between electric and magnetic modes, and in hot plasma both contribute at the same order. A comparative study between high temperature and zero temperature $\eta$ can be found in [22].

Another important quantity to study the equilibration property of the plasma is the momentum diffusion coefficient $B_{ij}$. The coefficient is related to the interaction rate *via* the following relation [24]:

$$B_{ij} = \int d\Gamma q_i q_j. \tag{24}$$

Decomposing $B_{ij}$ into longitudinal ($B_l$) and transverse components ($B_t$) we get the following expression,

$$B_{ij} = B_t \left( \delta_{ij} - \frac{p_i p_j}{p^2} \right) + B_l \frac{p_i p_j}{p^2}. \tag{25}$$

$B_{l,t}$ are the longitudinal, transverse squared momenta acquired by the particle through collision with the plasma. The longitudinal momentum diffusion coefficient is related to the drag coefficient *via* the Einstein relation in case of equilibrating plasma. Like $\eta$ longitudinal momentum diffusion coefficient ($B_l = \mathcal{B}$, suppressing the index $l$) can be expressed as

$$
\begin{aligned}
\mathcal{B} = \frac{\pi e^2}{E} & \int \frac{d^3 q}{(2\pi)^3} \int_{-\infty}^{\infty} \frac{dk_0}{2\pi} \rho_f(k_0) \int_{-\infty}^{\infty} \frac{dq_0}{2\pi} q_{\parallel}^2 \\
& \times (1 + n(q_0) - \bar{n}(k_0)) \, \delta(E - k_0 - q_0) \\
& \times [p_0 k_0 + \mathbf{p} \cdot \mathbf{k}] \, \rho_l(q_0, q) \\
& + 2 [p_0 k_0 - (\mathbf{p} \cdot \hat{\mathbf{q}})(\mathbf{k} \cdot \hat{\mathbf{q}})] \, \rho_t(q_0, q).
\end{aligned}
\tag{26}
$$

Here, $q_{\parallel}$ is the longitudinal exchanged momentum transfer in a scattering and $q_{\parallel} = q \cos \theta$. Explicitly $\mathcal{B}$ is then given by [2]

$$
\mathcal{B} \simeq \mathcal{B}_1 v^3 + \mathcal{B}_2 v^4 + \cdots,
\tag{27}
$$

where $\mathcal{B}_1 = (e^2 m_D^3)/(72\pi)$ and $\mathcal{B}_2 = (e^2 m_D^3)/128$. In this case also the electric and magnetic modes split as the magnetic mode is $\sim (E-\mu)^3$ whereas electric mode is $\sim (E-\mu)^4$. One can relate the leading order terms of $\eta$ and $\mathcal{B}$ from (22) and (27) via a common relation, namely, Einstein's relation as in case of high temperature plasma. In ultradegenerate plasma the relation can be written as $\mathcal{B} = 3E(E-\mu)\eta/4$.

In the high temperature and zero chemical potential regime $\mathcal{B}$ has the following expression [25–29]:

$$
\mathcal{B} \simeq \frac{e^4 T^3}{18\pi} \left( -\frac{1}{2} + \frac{3}{2} \log \left| \frac{q_{\max}}{m_D} \right| + \frac{1}{2} \log \left| \frac{2}{\sqrt{\pi}} \right| \right).
\tag{28}
$$

In case of high temperature plasma from (23) and (28), Einstein's relation becomes $\mathcal{B} = 2TE\eta$.

From (22), (23), (27), and (28) it can be seen that zero temperature $\eta$ and $\mathcal{B}$ show entirely different characteristics from the high temperature case. The logarithmic dependence in high temperature takes an algebraic form in the ultradegenerate limit [22].

### 3.2. Low Temperature Plasma.
In this section we extend our calculations to include finite temperature effects in the domain where $|E - \mu| \sim T \ll e\mu \ll \mu$. Moreover, we also go beyond the leading order calculations as previously reported above. Finally we shall see how the leading order zero temperature results appear as a limiting case from our present analysis including higher order correction terms.

For the soft sector contribution we take $T/m_D$ as an expansion parameter and $\alpha = (E - \mu)/T \sim 1$. We start with the following expression for the soft sector [4]:

$$
\begin{aligned}
\eta\big|^{\text{soft}}(E) \\
\simeq \frac{e^2}{8\pi^2 E} \int_0^{q^*} dq\, q^3 \int_{-1}^{1} dx\, x \left( 1 + n(qx) - \bar{n}(E - \mu - qx) \right) \\
\times \left\{ \rho_l(qx, q) + \left( 1 - x^2 \right) \rho_t(qx, q) \right\}.
\end{aligned}
\tag{29}
$$

The previous equation can easily be derived from (16). After subtracting the energy independent part from the above equation we get,

$$
\begin{aligned}
\eta\big|^{\text{soft}}(E) - \eta\big|^{\text{soft}}_{E=\mu} \\
= -\frac{e^2}{8\pi^2 E} \int_0^{q^*} dq\, q^3 \int_{-1}^{1} dx\, x \left( \bar{n}(E - \mu - qx) - \bar{n}(-qx) \right) \\
\times \left[ \left( 1 - x^2 \right) \rho_t(qx, q) + \rho_l(qx, q) \right].
\end{aligned}
\tag{30}
$$

We substitute $q$ and $q_0$ by introducing dimensionless variables $z$ and $\nu$ in the above equation:

$$
q = \frac{2 q_s z}{(\pi \nu)^{1/3}}, \qquad q_0 = T\nu,
\tag{31}
$$

where $q_s$ is the screening length in the magnetic sector and given by $(\pi m_D^2 q_0/2)^{1/3}$. This fractional power in the transverse sector will eventually show up in different coefficients as we show below. The previous substitutions actually yields,

$$
q = m_D a^{1/3} z, \qquad x = \frac{a^{2/3} \nu}{z}.
\tag{32}
$$

After performing the $q_0$ and $q$ integrations we obtain [4]:

$$
\begin{aligned}
\eta\big|^{\text{soft}}_t(E) - \eta\big|^{\text{soft}}_{t, E=\mu} \\
= \frac{e^2 m_D^2}{E} \\
\times \left\{ \frac{1}{48\pi} \left( \frac{T}{m_D} h_1 \left( \frac{(E-\mu)}{T} \right) \right)^2 \right. \\
- \frac{3 \times 2^{1/3}}{72\pi^{7/3}} \left( \frac{T}{m_D} h_2 \left( \frac{(E-\mu)}{T} \right) \right)^{8/3} \\
\left. - \frac{6 \times 2^{2/3}}{9\pi^{11/3}} \left( \frac{T}{m_D} h_3 \left( \frac{(E-\mu)}{T} \right) \right)^{10/3} \right\},
\end{aligned}
\tag{33}
$$

where

$$
\begin{aligned}
h_1 \left( \frac{(E-\mu)}{T} \right) &= \left[ \Gamma(3) \left( \text{Li}_2(-e^{-\alpha}) - \text{Li}_2(-e^{\alpha}) \right) \right]^{1/2}, \\
h_2 \left( \frac{(E-\mu)}{T} \right) &= \left[ \Gamma \left( \frac{11}{3} \right) \left( \text{Li}_{8/3}(-e^{-\alpha}) - \text{Li}_{8/3}(-e^{\alpha}) \right) \right]^{3/8}, \\
h_3 \left( \frac{(E-\mu)}{T} \right) &\left[ \Gamma \left( \frac{13}{3} \right) \left( \text{Li}_{10/3}(-e^{-\alpha}) - \text{Li}_{10/3}(-e^{\alpha}) \right) \right]^{3/10}.
\end{aligned}
\tag{34}
$$

In case of the longitudinal sector we substitute $q = q_s y$ and $q_0 = Tu/y$ or $q = m_D y$ and $x = au/y$. Since screening length is different in electric and magnetic sectors the substitutions therefore involve different coefficients of $m_D$ and $T$ for the transverse and the longitudinal cases [1] as can be seen from the structure of $\rho_{l,t}$ (8). For the leading term in the longitudinal sector one writes,

$$\eta|_l^{\text{soft}}(E) - \eta|_{l,E=\mu}^{\text{soft}} = \frac{e^2 m_D^2}{96 E}\left(\frac{T}{m_D} g_1\left(\frac{(E-\mu)}{T}\right)\right)^3, \quad (35)$$

where

$$g_1\left(\frac{(E-\mu)}{T}\right) = \left[\Gamma(4)\left(\text{Li}_3\left(-e^{-\alpha}\right) - \text{Li}_3\left(-e^{\alpha}\right)\right)\right]^{1/3}. \quad (36)$$

The final expression for drag coefficient then becomes [4]

$$\eta = \frac{e^2 m_D^2}{E}\left\{\frac{1}{48\pi}\left(\frac{T}{m_D} h_1\left(\frac{(E-\mu)}{T}\right)\right)^2\right.$$
$$- \frac{3\times 2^{1/3}}{72\pi^{7/3}}\left(\frac{T}{m_D} h_2\left(\frac{(E-\mu)}{T}\right)\right)^{8/3}$$
$$\left. - \frac{6\times 2^{2/3}}{9\pi^{11/3}}\left(\frac{T}{m_D} h_3\left(\frac{(E-\mu)}{T}\right)\right)^{10/3}\right\}$$
$$+ \frac{e^2 m_D^2}{96 E}\left(\frac{T}{m_D} g_1\left(\frac{(E-\mu)}{T}\right)\right)^3. \quad (37)$$

From the previous expression it is evident that $\eta$ is polylogarithmic in nature and contains fractional powers in $(E-\mu)$. The fractional powers indicate the nonanalytic behavior of $\eta$. Significant is to note that here the subleading transverse term is greater than the leading order longitudinal one. Thus we see that here too the transverse contribution dominates over the longitudinal one. In the zero temperature limit the functions behave as $h_i(\alpha) \to |\alpha|$ and $g_i(\alpha) \to |\alpha|$. Hence, $\eta$ in the limit becomes [4]

$$\eta = \mathscr{H}_1 \gamma^2 - \mathscr{H}_3 \gamma^{8/3} + \mathscr{H}_2 \gamma^3 + \cdots. \quad (38)$$

In the last equation $\mathscr{H}_3 = (3\times 2^{1/3} e^2 m_D^2)/(72\pi^{7/3} E)$, and $\mathscr{H}_1$, $\mathscr{H}_2$ are the same as before. We can infer from the previous observations that the inclusion of transverse interaction for a relativistic particle changes the nature of $\eta$. In the nonrelativistic case considering only the electric interaction $\eta$ goes as $(E-\mu)^3$, but with the inclusion of the transverse sector we see that anomalous fractional powers appear. Like $\eta$ one can also compute the longitudinal momentum diffusion coefficient in the low temperature limit. The expression for

the longitudinal momentum diffusion coefficient as defined in (26) can be written as [4]

$$\mathscr{B} = e^2 m_D^3\left\{\frac{1}{72\pi}\left(\frac{T}{m_D} h_4\left(\frac{(E-\mu)}{T}\right)\right)^3\right.$$
$$- \frac{3\times 2^{1/3}}{99\pi^{7/3}}\left(\frac{T}{m_D} h_5\left(\frac{(E-\mu)}{T}\right)\right)^{11/3}$$
$$\left. - \frac{20\times 2^{2/3}}{39\pi^{11/3}}\left(\frac{T}{m_D} h_6\left(\frac{(E-\mu)}{T}\right)\right)^{13/3}\right\}$$
$$+ \frac{e^2 m_D^3}{128}\left(\frac{T}{m_D} g_2\left(\frac{(E-\mu)}{T}\right)\right)^4, \quad (39)$$

where

$$h_4\left(\frac{(E-\mu)}{T}\right) = \left[\Gamma(4)\left(\text{Li}_3(-e^{-\alpha}) - \text{Li}_3(-e^{\alpha})\right)\right]^{1/3},$$
$$h_5\left(\frac{(E-\mu)}{T}\right) = \left[\Gamma\left(\frac{14}{3}\right)\left(\text{Li}_{11/3}(-e^{-\alpha}) - \text{Li}_{11/3}(-e^{\alpha})\right)\right]^{3/11},$$
$$h_6\left(\frac{(E-\mu)}{T}\right) = \left[\Gamma\left(\frac{16}{3}\right)\left(\text{Li}_{13/3}(-e^{-\alpha}) - \text{Li}_{13/3}(-e^{\alpha})\right)\right]^{3/13},$$
$$g_2\left(\frac{(E-\mu)}{T}\right) = \left[\Gamma(5)\left(\text{Li}_4\left(-e^{-\alpha}\right) - \text{Li}_4\left(-e^{\alpha}\right)\right)\right]^{1/4}. \quad (40)$$

The final expression for $\mathscr{B}$ in the extreme zero temperature limit becomes [4]

$$\mathscr{B} = \mathscr{B}_1 \gamma^3 - \mathscr{B}_3 \gamma^{11/3} + \mathscr{B}_2 \gamma^4 + \cdots, \quad (41)$$

$\mathscr{B}_3 = (2^{1/3} e^2 m_D^3)/(33\pi^{7/3})$, $\mathscr{B}_1$, and $\mathscr{B}_2$ are the same as earlier. It is to be noted that, for the low temperature case, so far, no reference has been made to the case of hard momentum regime. This is justified as we have seen before that up to the order of our interest the hard sector does not contribute, and the entire contribution comes from the soft sector alone.

Furthermore, it is worthwhile to mention that the NFL behaviour of these coefficients is related to the dynamical screening involving "$\omega$" in the static limit when one evaluates $\Pi_t$. To see this one can recall the expressions for the polarization functions in (6). In the static limit ($q_0/q \sim 0$) the functions become

$$\Pi_l = m_D^2, \qquad \Pi_t = \frac{i m_D^2 \pi\omega}{4q}. \quad (42)$$

The "$\omega$" that appears in the denominator of (5) from the $\Pi_t$ above along with the Pauli blocking is responsible for this NFL behavior.

In Figures 3 and 4 $\eta$ and $\mathscr{B}$ versus energy of the incoming fermion in the small temperature ($T/T_f \ll 1$) region ($T_f = \mu/k_B$ is the Fermi temperature) have been plotted. From the figures, it is evident that with increasing $T/T_f$, both $\eta$ and $\mathscr{B}$ decrease. This nature is consistent with what one finds for the fermionic damping rate at small temperature [1].

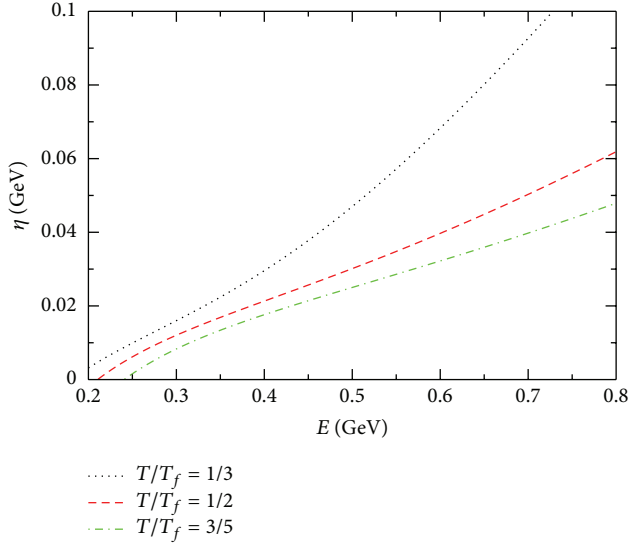

Figure 3: Energy dependence of the drag coefficient at $T/T_f = 1/3$ (dotted curve), $T/T_f = 1/2$ (dashed curve), and $T/T_f = 3/5$ (dash dotted curve) [10].

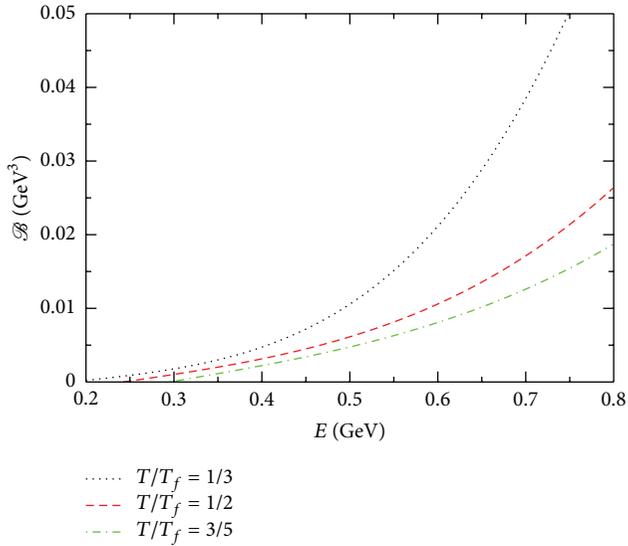

Figure 4: Energy dependence of the diffusion coefficient at $T/T_f = 1/3$ (dotted curve), $T/T_f = 1/2$ (dashed curve), and $T/T_f = 3/5$ (dash dotted curve) [10].

## 4. Neutrino Mean Free Path

In this section we calculate the neutrino mean free path (MFP) for cold and warm QCD matters. When a new star is born following a supernova explosion, it starts emitting neutrinos *via* the direct or the modified Urca processes. Emission of these neutrinos is responsible for the initial cooling of the star so produced. For quark matter, the dominant contribution to the emission of neutrinos is given by the $\beta$ decay and the electron capture [9]. These reactions are named as "quark direct Urca" processes which have been studied in detail by Iwamoto [30, 31]. For quark matter, the

MFP was previously derived in [30] where the calculations were restricted to the leading order by assuming free Fermi gas interactions. In this context we recall two important works on the neutrino MFP in QED plasma. One is due to Tubbs and Schramm [32], and the other is done by Lamb and Pethick [33]. In [32], MFP was calculated in the neutralized core and just outside the core. On the other hand, it has been shown in [33] that neutrino degeneracy reduces the neutrino MFP which suggests that neutrino may flow out of the core rather slowly. Here, we revisit the calculation and go beyond the leading order results to incorporate the NFL corrections both for degenerate and nondegenerate neutrinos.

In the interior of a neutron star, there are two distinct phenomena for which the neutrino mean free path is calculated: one is absorption, and the other involves scattering of neutrinos [31]. The corresponding mean free paths are denoted as $l_{\text{mean}}^{\text{abs}}$ and $l_{\text{mean}}^{\text{scatt}}$, and combine them to obtain the total mean free path [34]:

$$\frac{1}{l_{\text{mean}}^{\text{total}}} = \frac{1}{l_{\text{mean}}^{\text{abs}}} + \frac{1}{l_{\text{mean}}^{\text{scatt}}}. \tag{43}$$

In our model the Lagrangian density is described by [30]

$$\mathcal{L}_{Wx}(x) = \frac{G}{\sqrt{2}} l_\mu(x) \, \mathcal{J}_W^\mu(x) + \text{H.C.}, \tag{44}$$

where the weak coupling constant is $G \simeq 1.166 \times 10^{-11}$ in MeV units, and $l_\mu$ and $\mathcal{J}_W^\mu$ are the lepton and hadron charged weak currents, respectively. The weak currents are

$$l_\mu(x) = \bar{e}\gamma_\mu(1-\gamma_5)\nu_e + \bar{\mu}\gamma_\mu(1-\gamma_5)\nu_\mu + \cdots,$$

$$\mathcal{J}_W^\mu(x) = \cos\theta_c \bar{u}\gamma^\mu(1-\gamma_5)d + \sin\theta_c \bar{u}\gamma^\mu(1-\gamma_5)s + \cdots, \tag{45}$$

where $\theta_c$ is the Cabibbo angle ($\cos^2\theta_c \simeq 0.948$) [35].

The mean free path is determined by the quark neutrino interaction in dense quark matter *via* weak processes. We consider the simplest $\beta$ decay reactions: the absorption process

$$d + \nu_e \longrightarrow u + e^- \tag{46}$$

and the other is its inverse relation

$$u + e^- \longrightarrow d + \nu_e \tag{47}$$

The neutrino mean free path is related to the total interaction rate due to neutrino emission averaged over the initial quark spins and summed over the final state phase space and spins. It is given by [31]

$$\frac{1}{l_{\text{mean}}^{\text{abs}}(E_\nu, T)} = \frac{g}{2E_\nu} \int \frac{d^3 p_d}{(2\mathbf{p})^3} \frac{1}{2E_d} \int \frac{d^3 p_u}{(2\mathbf{p})^3} \frac{1}{2E_u}$$

$$\times \int \frac{d^3 p_e}{(2\mathbf{p})^3} \frac{1}{2E_e} (2\pi)^4 \delta^4 \left(P_d + P_\nu - P_u - P_e\right),$$

$$|M|^2 \{ n(p_d) [1 - n(p_u)] [1 - n(p_e)]$$

$$- n(p_u) n(p_e) [1 - n(p_d)] \}, \tag{48}$$

where $g$ is the spin and color degeneracy, which in the present case is considered to be 6. Here, $E$, $p$, and $n_p$ are the energy, momentum, and distribution function for the corresponding particle. $|M|^2$ is the squared invariant amplitude averaged over initial $d$ quark spin and summed over final spins of $u$ quark and electron as given by [31]

$$|M|^2 = \frac{1}{2} \sum_{\sigma_u, \sigma_d, \sigma_e} \left|M_{fi}\right|^2 = 64 G^2 \cos^2\theta_c \left(P_d \cdot P_\nu\right) \left(P_u \cdot P_e\right). \tag{49}$$

Here, we work with the two flavor system as the interaction involving strange quark is Cabibbo suppressed [9, 36].

### 4.1. Degenerate Neutrinos.
In this section we consider the case of degenerate neutrinos. When the neutrino chemical potential $(\mu_\nu)$ is considered to be much larger than the temperature $(T)$, the neutrinos become degenerate. This is also termed as trapped neutrino matter. In this case both the direct (46) and inverse (47) processes can occur, and both terms in (48) under curly brackets are retained [31]. Consequently, the $\beta$ equilibrium condition becomes $\mu_d + \mu_\nu = \mu_u + \mu_e$. Neglecting the quark-quark interactions and by using (48) and (49), the mean free path becomes

$$\begin{aligned}
\frac{1}{l_{\text{mean}}^{\text{abs},D}} &= \frac{3}{4\pi^5} G^2 \cos^2\theta_c \int d^3 P_d \int d^3 P_u \\
&\quad \times \int d^3 P_e \left(1 - \cos\theta_{d\nu}\right) \left(1 - \cos\theta_{ue}\right) \\
&\quad \times \delta^4 \left(P_d + P_\nu - P_u - P_e\right) \left[1 + e^{-\beta(E_\nu - \mu_\nu)}\right] \\
&\quad \times n\left(p_d\right) \left[1 - n\left(p_u\right)\right] \left[1 - n\left(p_e\right)\right].
\end{aligned} \tag{50}$$

In the square bracket, the second term $e^{-\beta(E_\nu - \mu_\nu)}$ is due to the inverse process (47). One can neglect the mass effect on the neutrino MFP as the masses of $u$, $d$ quark and electron are very small. To perform the momentum integration we define $p \equiv |p_d + p_\nu| = |p_u + p_e|$ as a variable. Following the same procedure as described by Iwamoto [31] one has

$$\sin\theta_{d\nu} d\theta_{d\nu} = \frac{p \, dp}{p_f(d) \, p_f(\nu)}, \tag{51}$$

$$\left(1 - \cos\theta_{d\nu}\right) \left(1 - \cos\theta_{ue}\right) \simeq \frac{p^4 - 2p^2 p_f^2 + p_f^4}{4 p_f(d) \, p_f(\nu) \, p_f(u) \, p_f(e)}, \tag{52}$$

$$\begin{aligned}
d^3 P_d &= 2\pi \sin\theta_{d\nu} d\theta_{d\nu} p_d^2 dp_d \\
&= 2\pi \frac{p_f(d)}{p_f(\nu)} p \, dp \frac{dp_d}{dE_d} dE_d \\
&= 2\pi \frac{p_f(d)}{p_f(\nu)} p \, dp \frac{dp_d}{d\omega} d\omega,
\end{aligned} \tag{53}$$

$$d^3 P_u = 2\pi \frac{p_f(u) \, p_f(e)}{p} dE_e \frac{dp_u}{d\omega} d\omega, \tag{54}$$

where we denote the single particle energy $E_{d(u)}$ as $\omega$. For the free case $dp/d\omega$ is the inverse quark velocity. It is well known that this slope of the dispersion relation changes in matter due to the scattering from the Fermi surface and excitation of the Dirac vacuum. The modified dispersion relation can be obtained by computing the on-shell one-loop self-energy. For quasiparticles with momenta close to the Fermi momentum, the one-loop self-energy is dominated by the soft gluon exchanges [8]. The quasiparticle energy $\omega$ satisfies the relation [8, 21]

$$\omega = E_p(\omega) - \text{Re}\,\Sigma\left(\omega, p(\omega)\right), \tag{55}$$

where we have retained only the real part of self-energy since the imaginary part of $\Sigma$ turns out to be negligible compared to its real part [1, 37]. For the detailed analysis we refer the reader to [11, 12, 21].

The analytical expressions for one-loop quark self-energy can be written as [1, 8, 37–40]

$$\Sigma = \frac{g^2 C_F}{12\pi^2} \left(\omega - \mu\right) \ln\left(\frac{m_D}{\omega - \mu}\right) + i \frac{g^2 C_F}{24\pi} |\omega - \mu|. \tag{56}$$

It exhibits a logarithmic singularity close to the Fermi surface, that is, when $\omega \to \mu$.

$dp(\omega)/d\omega$ can be obtained by differentiating the dispersion relation (55) with respect to $p$. At leading order in $T/\mu$ this yields

$$\begin{aligned}
\frac{dp(\omega)}{d\omega} &\simeq \left(1 - \frac{\partial}{\partial\omega} \text{Re}\,\Sigma(\omega)\right) \frac{E_p(\omega)}{p(\omega)} \\
&= \left[1 + \frac{C_F \alpha_s}{3\pi} \ln\left(\frac{m_D}{T}\right)\right] \frac{E_p(\omega)}{p(\omega)},
\end{aligned} \tag{57}$$

where $\alpha_s$ is the strong coupling constant, $C_F = (N_c^2 - 1)/(2N_c)$, and $N_c$ is the color factor. Using (57), (50), and (52)–(54), the neutrino mean free path can be determined for two conditions. For $|p_f(u) - p_f(e)| \geq |p_f(d) - p_f(\nu)|$

$$\begin{aligned}
\frac{1}{l_{\text{mean}}^{\text{abs},D}} &= \frac{4}{\pi^3} G^2 \cos^2\theta_c \frac{\mu_u^2 \mu_e^2}{\mu_\nu^2} \left[1 + \frac{1}{2}\left(\frac{\mu_e}{\mu_u}\right) + \frac{1}{10}\left(\frac{\mu_e}{\mu_u}\right)^2\right] \\
&\quad \times \left[\left(E_\nu - \mu_\nu\right)^2 + \pi^2 T^2\right] \left[1 + \frac{C_F \alpha_s}{3\pi} \ln\left(\frac{m_D}{T}\right)\right]^2.
\end{aligned} \tag{58}$$

To derive (58), we use the chemical equilibrium condition $p_f(u) + p_f(e) = p_f(d) + p_f(\nu)$ neglecting the masses for quarks and electrons. For the phase-space integral we have

$$\begin{aligned}
\int_0^\infty dE_d &\int_0^\infty dE_u \\
&\times \int_0^\infty dE_e \left[1 + e^{-\beta(E_\nu - \mu_\nu)}\right] n\left(p_d\right) \\
&\times \left[1 - n\left(p_u\right)\right] \left[1 - n\left(p_e\right)\right] \delta\left(E_d + E_\nu - E_u - E_e\right) \\
&\simeq \frac{1}{2} \left[\left(E_\nu - \mu_\nu\right)^2 + \pi^2 T^2\right].
\end{aligned} \tag{59}$$

Similarly, for $|p_f(d) - p_f(\nu)| \geq |p_f(u) - p_f(e)|$, the corresponding expression for mean free path can be obtained by replacing $\mu_u \leftrightarrow \mu_d$ and $\mu_e \leftrightarrow \mu_\nu$ in (58).

The other contribution to the mean free path comes from the quark-neutrino scattering. The neutrino scattering process from degenerate quarks is given by

$$q_i + \nu_e\,(\bar{\nu}_e) \longrightarrow q_i + \nu_e\,(\bar{\nu}_e) \tag{60}$$

for each quark component of flavor $i$ (= $u$ or $d$). The scattering mean free path of the neutrinos in degenerate case can be calculated similarly as evaluated by Lamb and Pethick in [33] for electron-neutrino scattering. Assuming $m_{q_i}/p_{f_i} \ll 1$ and including the suitable modifications of the phase space, the mean free path is given by

$$\frac{1}{l_{\text{mean}}^{\text{scatt},D}} = \frac{3}{16} n_{q_i} \sigma_0 \left[ \frac{\left(E_\nu - \mu_\nu\right)^2 + \pi^2 T^2}{m_{q_i}^2} \right]$$
$$\times \left[ 1 + \frac{C_F \alpha_s}{3\pi} \ln\left(\frac{m_D}{T}\right) \right]^2 \Lambda\,(x_i). \tag{61}$$

Here, $m_{q_i}$ is the quark mass. $C_{V_i}$ and $C_{A_i}$ are the vector and axial vector coupling constant given in Table II of [31]. The expression of (61) has been found to be in agreement with the results reported in [33] for dense and cold QED plasmas by making suitable changes for the color factors and by dropping the second square bracketed term. In (61) the constants $\sigma_0 \equiv 4G^2 m_{q_i}^2/\pi$ [32], and $n_{q_i}$ is the quark number density:

$$n_{q_i} = 2 \int \frac{d^3 p}{(2\mathbf{p})^3} \frac{1}{e^{\beta(E_{q_i} - \mu_{q_i})}}, \tag{62}$$

where 2 is the quark spin degeneracy factor. Explicit form of $\Lambda(x_i)$ can be written as [31, 33]

$$\Lambda\,(x_i) = \frac{4}{3} \frac{\text{Min}\left(\mu_\nu, \mu_{q_i}\right)}{\mu_{q_i}}$$
$$\times \left[ \left(C_{V_i}^2 + C_{A_i}^2\right)\left(2 + \frac{1}{5}x_i^2\right) + 2C_{V_i}C_{A_i}x_i \right], \tag{63}$$

and $x_i = \mu_\nu/\mu_{q_i}$ if $\mu_\nu < \mu_{q_i}$, and $x_i = \mu_{q_i}/\mu_\nu$ if $\mu_\nu > \mu_{q_i}$.

### 4.2. Nondegenerate Neutrinos.

The mean free path has been also derived for nondegenerate neutrinos, that is, when $\mu_\nu \ll T$. This is the case for untrapped neutrino matter. For this case the inverse process (47) does not contribute to the MFP. Hence we neglect the second term in the curly braces of (48). Here, only those fermions take part in the reaction whose momenta lie close to their respective Fermi surfaces. It is to be mentioned here, if quarks are treated as free, as discussed in [31, 42, 43], the matrix element vanishes since $u$, $d$ quarks and electrons are collinear in momenta. The inclusion of strong interactions between quarks relaxes these kinematic restrictions resulting in a nonvanishing squared matrix amplitude. Since the neutrinos are produced thermally, we neglect the neutrino momentum in energy-momentum conservation relation [31]. This is not the case for degenerate neutrinos where $p_\nu \gg T$, and therefore such approximation is not valid there. By doing angular average over the direction of the outgoing neutrino, from (49) the squared matrix element is given by [9]:

$$|M|^2 = 64G^2\cos^2\theta_c\, p_f^2\,(V_d \cdot P_\nu)\,(V_u \cdot P_e)$$
$$= 64G^2\cos^2\theta_c\, p_f^2\, E_\nu \mu_e \frac{C_F \alpha_s}{\pi}, \tag{64}$$

where $V = (1, \nu_f)$ is the four velocity. To calculate (64) we have used the chemical equilibrium condition $\mu_d = \mu_u + \mu_e$ and also the relations derived from Fermi liquid theory [9]

$$\nu_F = 1 - \frac{C_F \alpha_s}{2\pi}; \qquad \delta\mu = \frac{C_F \alpha_s}{\pi}\mu. \tag{65}$$

Putting $|M|^2$ in (48) we have

$$\frac{1}{l_{\text{mean}}^{\text{abs},ND}} = \frac{3C_F \alpha_s}{4\mathbf{p}^6}G^2\cos^2\theta_c \int d^3 p_d \int d^3 p_u \int d^3 p_e$$
$$\delta^4\left(P_d + P_\nu - P_u - P_e\right)n\,(p_d)\left[1 - n\,(p_u)\right]\left[1 - n\,(p_e)\right]. \tag{66}$$

Neglecting the neutrino momentum in the neutrino momentum conserving $\delta$ function, the integrals can be decoupled into two parts. Following the procedure described by Iwamoto [31], the angular integral is given by

$$\mathscr{A} = \int d\Omega_d \int d\Omega_u \int d\Omega_e \delta\,(p_d - p_u - p_e) = \frac{8\pi^2}{\mu_d \mu_u \mu_e} \tag{67}$$

and the other part

$$\mathscr{D} = \int_0^\infty p_d^2 \frac{dp_d}{dE_d}dE_d \int_0^\infty p_u^2 \frac{dp_u}{dE_u}dE_u \int_0^\infty p_e^2\,dE_e$$
$$\delta\left(E_d + E_\nu - E_u - E_e\right)n\,(p_d)\left[1 - n\,(p_u)\right]\left[1 - n\,(p_e)\right]. \tag{68}$$

Changing the variables to $x_d = (E_d - \mu_d)\beta$, $x_u = -(E_u - \mu_u)\beta$, and $x_e = -(E_e - \mu_e)\beta$ and denoting the single particle energy $E_{u(d)}$ as $\omega$ we have from (68)

$$\mathscr{D} = \int_{-\infty}^\infty dx_d dx_u dx_e \frac{dp_d\,(\omega)}{d\omega}\frac{dp_u\,(\omega)}{d\omega}p_d^2 p_u^2 p_e^2 \delta$$
$$\times\,(x_d + x_u + x_e + \beta E_\nu)n\,(x_d)\,n\,(-x_u)\,n\,(-x_e). \tag{69}$$

As the contribution dominates near the Fermi surfaces, extension of lower limit is reasonable [41, 44].

The integration of (69) can be performed using (57) and following the procedure defined in [37, 41, 44] to give

$$\mathscr{D} = \mu_d^2 \mu_u^2 \mu_e^2 \frac{\left(E_\nu^2 + \pi^2 T^2\right)}{2\left(1 + e^{-\beta E_\nu}\right)}\left[1 + \frac{C_F \alpha_s}{3\pi}\ln\left(\frac{m_D}{T}\right)\right]^2. \tag{70}$$

Using (66), (67), and (70), the mean free path at leading order in $T/\mu$ is given by

$$\frac{1}{l_{\text{mean}}^{\text{abs},ND}} = \frac{3C_F \alpha_s}{\pi^4}G^2\cos^2\theta_c \mu_d \mu_u \mu_e$$
$$\times \frac{\left(E_\nu^2 + \pi^2 T^2\right)}{\left(1 + e^{-\beta E_\nu}\right)}\left[1 + \frac{C_F \alpha_s}{3\pi}\ln\left(\frac{m_D}{T}\right)\right]^2. \tag{71}$$

The first term is known from [31], and the additional terms are higher order corrections to the previous results derived in the present work.

For the scattering of nondegenerate neutrinos in quark matter, the expression of mean free path was given by Iwamoto [31]. We incorporate the anomalous effect which enters through phase space modification giving rise to

$$
\frac{1}{l_{\text{mean}}^{\text{scatt},ND}} = \frac{C_{V_i}^2 + C_{A_i}^2}{20} n_{q_i} \sigma_0
$$
$$
\times \left(\frac{E_\nu}{m_{q_i}}\right)^2 \left(\frac{E_\nu}{\mu_i}\right) \left[1 + \frac{C_F \alpha_s}{3\pi} \ln\left(\frac{m_D}{T}\right)\right]^2. \tag{72}
$$

Here, we have assumed $m_{q_i}/p_{f_i} \ll 1$. The constants $\sigma_0$ and number density $n_{q_i}$ have been defined earlier.

We now estimate the numerical values of the neutrino mean free paths. Here, $E_\nu$ is set to be equal to $3T$ and $m_q = 10\,$MeV [31, 34]. For the quark chemical potential, following [9], we take $\mu_q \simeq 500\,$MeV corresponding to densities $\rho_b \approx 6\rho_0$, where $\rho_0$ is the nuclear matter saturation density. The electron chemical potential is determined by using the charge neutrality and beta equilibrium conditions which yields $\mu_e = 11\,$Mev. The other parameters used are the same as [9].

From Figure 5, we find for degenerate neutrinos at the anomalous logarithmic terms reduce the value of the mean free path appreciably both in the low and high temperature regimes. Figure 6 shows that for nondegenerate neutrinos NFL correction is quite large at low temperature while at higher temperature it tends to merge. It is interesting to see from these two plots that NFL correction to the MFP in degenerate neutrinos is less than that of nondegenerate neutrinos [3]. This reduced mean free path is expected to influence the cooling of the compact stars.

## 5. Cooling of Neutron Star

To analyze the cooling of the compact star [45–47], the specific heat capacity of the quark matter core needs to be taken into consideration along with the emissivity via the cooling equation [9, 36]

$$
\frac{\partial u}{\partial t} = \frac{\partial u}{\partial T}\frac{\partial T}{\partial t} = C_\nu(T)\frac{dT}{dt} = -\epsilon(T), \tag{73}
$$

where $u$ is the internal energy, $C_\nu$ is the specific heat at constant volume and baryon number, $\epsilon$ is the emissivity, $t$ is time, and $T$ is the temperature, and we have assumed that there is no surface emission.

Emissivity under different scenarios both for quark and nuclear matters has been studied extensively. For the specific heat however earlier literatures were concerned only with the kinetic energy contribution without paying much attention to the QCD interaction. Recently several authors have shown that the interaction can lead to a significant correction to $C_\nu$. It has been shown that this interaction mediated by the gluons can give rise to a leading $T \ln T^{-1}$ term in the specific heat [48, 49]. This logarithmic enhancement comes from long-range chromomagnetic fields which eventually leads to non-Fermi-liquid behavior. The calculation of $C_\nu$ including NFL

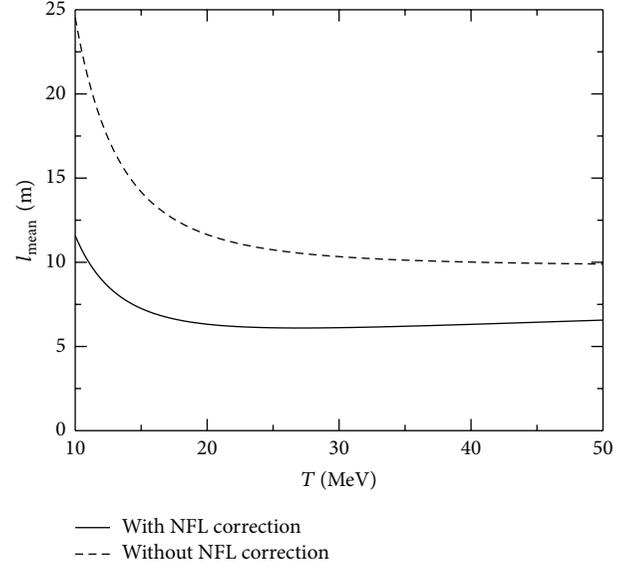

Figure 5: Neutrino mean free path in quark matter for degenerate neutrinos [3].

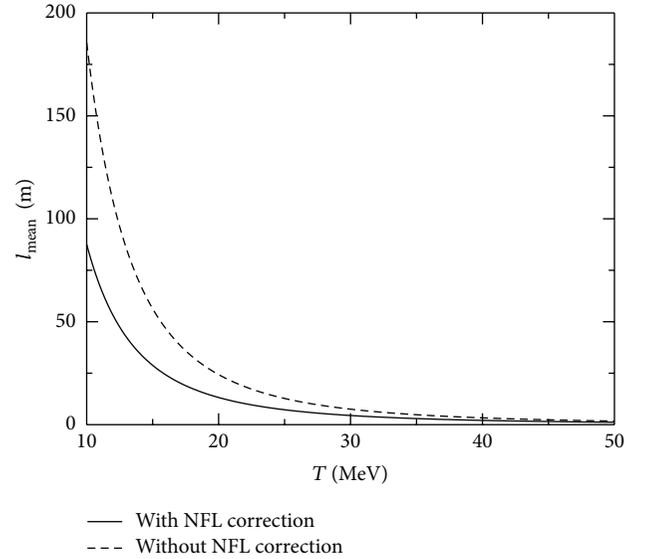

Figure 6: Neutrino mean free path in quark matter for nondegenerate neutrinos [3].

correction has been done in [48, 49] which we quote here for further use:

$$
C_\nu = \frac{N_c N_f}{3}\mu^2 T + \frac{N_c N_f}{9\pi}\mu^2 C_F \alpha_s T \ln\left(\frac{0.28 m_D}{T}\right). \tag{74}
$$

On the other hand, the neutrino emissivity of degenerate quark matter was first derived by Iwamoto [31]. Latter Schäfer and Schwenzer [9] calculated the neutrino emissivity including the corrections for NFL effects. The calculation has been made by considering the energy loss due to neutrino emission from the core of the neutron star. Alternatively, one can also determine the neutrino emissivity from the mean free path by evaluating the integral [31]

$$
\begin{aligned}
\epsilon &= \int \frac{d^3 p_\nu}{(2\mathbf{p})^3} \frac{E_\nu}{I_{\text{mean}}^{\text{abs},ND}(-E_\nu, T)} \\
&\approx \frac{457}{630} G^2 \cos^2\theta_c \alpha_s \mu_u \mu_d \mu_e T^6 \left(1 + \frac{C_F \alpha_s}{3\pi} \log\left(\frac{m_D}{T}\right)\right)^2.
\end{aligned}
\tag{75}
$$

This is the same as given in [9]. In this context, it is to be mentioned that without such non-Fermi liquid effects we have $\epsilon \sim T^6$ and $C_\nu \sim T$. In this case the temperature scales turn out to be $T \propto 1/t^{1/4}$ [9]. With logarithmic corrections included, no simple analytic solution is possible, and we solve (73) numerically.

We observe from Figure 7 that the cooling of neutron star is significantly faster with NFL corrections as compared to the Fermi liquid result [9].

## 6. Thermal Relaxation Time

In this section we address the issue of thermal conduction that follows the initial cooling by the neutrino emission. This phenomenon is intimately connected to the thermal relaxation mechanism. A new born star emits a large amount of neutrinos to cool the core, but the crust remains hot. Hence, a temperature gradient is set up between the crust and the core. Immediately after the neutrino emission discussed earlier in the paper thermal energy gradually flows from outer crust to the inner core by heat conduction which, alternatively, might be viewed as the propagation of the cooling waves from the center towards the surface leading to thermalization. We in this section estimate the thermalization time scale ($\tau_\kappa$) and reveal the NFL behavior of the quantity in the context of the neutron star crust. In the latter case the electrons constitute an almost ideal Fermi gas and scatter between themselves. The kinetic theory definition of $\tau_\kappa$ is

$$
\tau_\kappa = \frac{3\kappa}{C_\nu}.
\tag{76}
$$

In case of the strongly degenerate electron gas the electron thermal conductivity ($\kappa_e$) can be expressed in terms of the thermal current $J_T$ as follows:

$$
\kappa_e = \frac{J_T}{T \nu_e}, \quad \nu_e = \nu_{ei} + \nu_{ee},
\tag{77}
$$

where $\nu_e$ is the total effective collision frequency. In the neutron star crust the main components that contribute to different transport coefficients are electrons and ions. Hence, the total collision frequency is the sum of the partial collision frequencies of electron-ion ($\nu_{ei}$) and electron-electron ($\nu_{ee}$). Evidently $\kappa_e$ is related to $\kappa_{ee}$ and $\kappa_{ei}$ via the following expression [50, 51]:

$$
\frac{1}{\kappa_e} = \frac{1}{\kappa_{ei}} + \frac{1}{\kappa_{ee}},
$$
$$
\kappa_{ei} = \frac{J_T}{T \nu_{ei}}, \quad \kappa_{ee} = \frac{J_T}{T \nu_{ee}}.
\tag{78}
$$

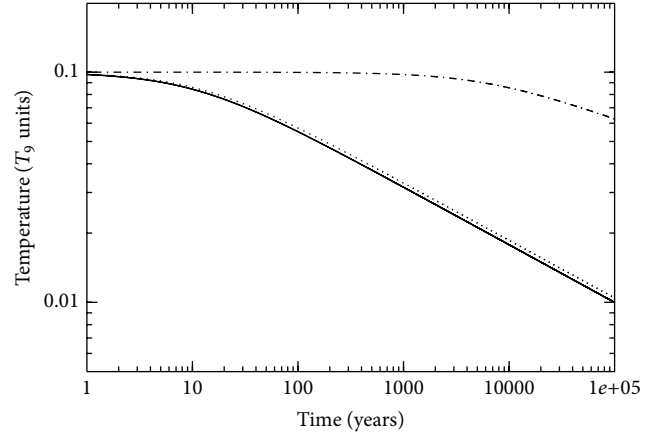

Figure 7: The cooling behavior of neutron star with core as neutron matter and degenerate quark matter. The dotted line represents the Fermi liquid result; the solid line gives the result including non-Fermi liquid correction. The dash-dotted line gives the cooling behavior of the neutron star core made up of purely neutron matter.

Thus the derivation of $\kappa_{ee}$ involves the calculation of $\nu_{ee}$. To derive $\nu_{ee}$ we start with the Boltzmann equation which describes the kinetics of the particles:

$$
\left(\frac{\partial}{\partial t} + \mathbf{v_p} \cdot \nabla_r + \mathbf{F} \cdot \nabla_p\right) n_p = -\mathscr{C}[n_p],
\tag{79}
$$

where $\mathbf{p}$ is the momentum of the quasiparticle, $\mathbf{F}$ is the external force, $\mathbf{v_p}$ is the velocity of the heat carrier, and $n_p$ is the distribution function of electrons. The right hand side of the Boltzmann equation is the collision integral describing the rate of scattering of fermions. In the presence of a weak stationary temperature gradient Boltzmann equation takes the following form:

$$
\mathbf{v_p} \cdot \nabla_r n_p = -\mathscr{C}[n_p].
\tag{80}
$$

The Fermi-Dirac distribution functions deviate from equilibrium distribution functions $n_i$ due to the presence of a weak temperature gradient, which we write as

$$
\widetilde{n}_i = n_i + \frac{\partial n_i}{\partial \epsilon_i} \Phi_i \frac{\nabla T}{T},
\tag{81}
$$

where

$$
n_i = \left\{\exp\left(\frac{\epsilon_i - \mu}{T}\right) + 1\right\}^{-1},
\tag{82}
$$

$\epsilon$ is the particle energy, $\mu$ is the chemical potential, and $T$ is the temperature. The term with $\Phi$ measures the deviation from equilibrium. The collision integral can be written as follows:

$$
\begin{aligned}
\mathscr{C}[n_p] = \nu\nu' \int_{p',k,k'} &\widetilde{n_p}\widetilde{n_k}\left(1 \pm \widetilde{n_p'}\right)\left(1 \pm \widetilde{n_k'}\right) \\
&- \widetilde{n_p'}\widetilde{n_k'}\left(1 \pm \widetilde{n_p}\right)\left(1 \pm \widetilde{n_k}\right)(2\pi)^4 \delta^4 \\
&\times \left(p + k - p' - k'\right)|M|^2.
\end{aligned}
\tag{83}
$$

In the previous equation $\nu$, $\nu'$ are the degeneracy factors and $|M|^2$ is the squared matrix element for the $2 \rightarrow 2$ scattering process. The $\pm$ sign is meant for stimulated emission or Pauli blocking. Since we consider only the electron-electron scattering, the negative sign will remain in the phase space factor. We write the previous equation in the form $|X\rangle = I|\Phi\rangle$, where $|X\rangle = (\epsilon_p - \mu)\nu_p$ and $I$ is the integral operator. The thermal conductivity $\kappa_{ee}$ is then given by the maximum of the following equation [52, 53]:

$$\kappa_{ee} = \frac{\langle X \mid \Phi \rangle^2}{T \langle \Phi |I| \Phi \rangle}; \tag{84}$$

$\langle \cdot \mid \cdot \rangle$ denotes an inner product; for $\Psi = \Phi$ the quantity $\langle X|\Psi\rangle^2/T\langle \Psi|I|\Psi\rangle$ is minimal with the minimal value $\kappa_{ee}$. With the help of this definition one can write

$$\frac{1}{\kappa_{ee}} \geq \left( \nu \int_p \frac{(\epsilon_p - \mu)}{T} \nu_z n_p \left( 1 - n_p \right) \Psi_p \right)^{-2}$$
$$\times \nu\nu' \int_{p,p',k,k'} n_p n_k \left( 1 - n'_p \right) \left( 1 - n'_k \right) \tag{85}$$
$$\times (2\pi)^4 \delta^4 \left( p + k - p' - k' \right) |M|^2$$
$$\times \frac{\left( \Psi_p + \Psi_k - \Psi_{p'} - \Psi_{k'} \right)^2}{4};$$

the first bracketed term in the denominator is the thermal current $J_T$. In principle $\Phi$ should be determined by minimizing (84) using the variational principle. But in the present scenario we consider the simplest trial function as [50, 52]

$$\Psi_p \propto \left( \epsilon_p - \mu \right) \nu_z. \tag{86}$$

The term in the bracket can be averaged over the $z$ axis keeping $x$ and $\phi$ fixed using the previous trial function, where, $\phi$ is the azimuthal angle between $\nu_p$ and $\nu_k$. After averaging we obtain [52]

$$\left( \Psi_p + \Psi_k - \Psi_{p'} - \Psi_{k'} \right)^2 = \frac{2}{3}\omega^2 \left( 1 - x^2 \right) (1 - \cos\phi). \tag{87}$$

For small energy transfer the electron-electron scattering squared matrix element is given by [52]

$$|M|^2 = 32e^4 \left[ \frac{1}{(q^2 + \Pi_l)} + \frac{(1 - x^2)\cos\phi}{(q^2 - \omega^2 + \Pi_t)} \right]^2. \tag{88}$$

The medium modified photon propagator in the previous equation contains the polarization functions $\Pi_l(q_0,q)$ and $\Pi_t(q_0,q)$, which describe plasma screening of interparticle interaction by longitudinal and transverse plasma perturbations, respectively. Now, we first analyze the denominator

of (85). The denominator is the thermal current as defined earlier and is given by

$$J_T = \nu \int_p \frac{(\epsilon_p - \mu)}{T} \nu_z n_p \left( 1 - n_p \right) \Psi_p$$
$$= \frac{\nu\mu^2 T^2}{6}. \tag{89}$$

Here, the degeneracy factor for electrons is $\nu = 2$.

The fermionic dispersion relation gets modified due to the inclusion of the fermion self-energy in the presence of the medium. Hence, energy momentum relation changes from the vacuum. For the momentum integration in the phase-space integral in (85) we will use the medium modified dispersion relation given in (55). For this one needs to know the fermion self-energy. The fermion self-energy for the QCD matter has already been quoted in (56). In case of electrons at low temperature with NLO correction it is given by the following [1]:

$$\Sigma(\epsilon, k)$$
$$= e^2 m$$
$$\times \left\{ \frac{\epsilon}{12\pi^2 m} \left[ \log\left( \frac{4\sqrt{2}m}{\pi\epsilon} \right) + 1 \right] \right.$$
$$+ \frac{i\epsilon}{24\pi m} + \frac{2^{1/3}\sqrt{3}}{45\pi^{7/3}} \left( \frac{\epsilon}{m} \right)^{5/3} \left( \text{sgn}(\epsilon) - \sqrt{3}i \right)$$
$$+ \frac{i}{64\sqrt{2}} \left( \frac{\epsilon}{m} \right)^2 - 20\frac{2^{2/3}\sqrt{3}}{189\pi^{11/3}} \left( \frac{\epsilon}{m} \right)^{7/3} \left( \text{sgn}(\epsilon) + \sqrt{3}i \right)$$
$$- \frac{6144 - 256\pi^2 + 36\pi^4 - 9\pi^6}{864\pi^6} \left( \frac{\epsilon}{m} \right)^3$$
$$\times \left[ \log\left( \frac{0.928m}{\epsilon} \right) - \frac{i\pi \, \text{sgn}(\epsilon)}{2} \right] + \mathcal{O}\left( \left( \frac{\epsilon}{m} \right)^{11/3} \right) \right\}, \tag{90}$$

where $m^2 = m_D^2/2$ and $\epsilon$ is chosen to be $(\epsilon_k - \mu)$. The phase-space correction due to the medium modified dispersion relation can now be written as [10]

$$\frac{dk}{d\epsilon_k} \simeq 1 + \frac{e^2}{12\pi^2} \log\left( \frac{4}{\pi\lambda} \right)$$
$$+ \frac{2^{1/3}e^2\lambda^{2/3}}{9\sqrt{3}\pi^{7/3}} - \frac{40 \times 2^{1/3}e^2\lambda^{4/3}}{27\sqrt{3}\pi^{11/3}} \cdots \tag{91}$$
$$= (1 + \beta),$$

where $\lambda = T/m_D$. The expression for $\kappa_{ee}^{-1}$ now becomes

$$\frac{1}{\kappa_{ee}} = \frac{\mathcal{P}}{T^2} \frac{dk}{d\epsilon_k} \frac{dp}{d\epsilon_p} I_{\kappa_{ee}} \left( \frac{T}{m_D} \right), \tag{92}$$

where $\mathscr{P} = 3e^4/4\pi^5$ and

$$
I_{\kappa_{ee}}\left(\frac{T}{m_D}\right) \equiv \int_0^\infty \frac{d\omega}{\omega} \frac{\omega^2/4T^2}{(\sinh{(\omega/2T)})^2}
$$
$$
\times \int_0^1 dx \int_0^{2\pi} \frac{d\phi}{2\pi} x^2 \left(1 - x^2\right)\left(1 - \cos\phi\right)
$$
$$
\times \left| \frac{1}{1 + (xm_D/\omega)^2 \chi_l(x)} \right.
$$
$$
\left. - \frac{\cos\phi}{1 + (xm_D/\omega)^2 \chi_t(x)/(1-x^2)} \right|^2. \tag{93}
$$

In the previous equation we have used the following expression [10, 52]:

$$
\int d\epsilon_p \, d\epsilon_k \; n_{\epsilon_p} n_{\epsilon_k} \left(1 - n_{(\epsilon_p + \omega)}\right)\left(1 - n_{(\epsilon_k - \omega)}\right)
$$
$$
= (1 + \beta)\, T^2 \frac{\omega^2/4T^2}{(\sinh{(\omega/2T)})^2}; \tag{94}
$$

while writing the previous equation we have considered the medium modified phase-space factor only for the particle with momentum $k$. One can in principle take the correction for both the particles with momentum $p$ and $k$, but that will contribute at higher order in coupling constant.

The final expression for the electron thermal conductivity now takes the following form [10]:

$$
\kappa_{ee} = \left[ \frac{\mathscr{P}}{T^2}\left(1 + \beta\right) \right.
$$
$$
\times \left\{ 2\lambda^2 \zeta\left(3\right) + \frac{(2\pi)^{2/3}}{3} \lambda^{8/3} \zeta\left(\frac{11}{3}\right) \Gamma\left(\frac{14}{3}\right) \right. \tag{95}
$$
$$
\left.\left. + \frac{\pi^5}{15}\lambda^3 \right\} \right]^{-1}.
$$

Unlike the Fermi-liquid result where $\kappa_{ee} \propto 1/T$, here the temperature dependence is nonanalytical and anomalous in nature. This is reminiscent of other quantities involving ultradegenerate plasma presented in the previous sections [1, 3–5]. $\kappa_{ee}$ involves fractional powers in $(T/m_D)$ coming from the medium modified phase space factor too.

For the estimation of relaxation time the other quantity which we require is the specific heat too. For the degenerate electron gas it can be written as [48, 49]

$$
C_v = \frac{\mu^2 T}{3} + \frac{m_D^2 T}{36}\left(\ln\left(\frac{4}{\pi\lambda}\right) + \gamma_E - \frac{6}{\pi^2}\zeta'\left(2\right) - 3\right)
$$
$$
- 40\frac{2^{2/3}\Gamma\left(8/3\right)\zeta\left(8/3\right)m_D^3}{27\sqrt{3}\pi^{7/3}}\lambda^{5/3} \tag{96}
$$
$$
+ 560\frac{2^{1/3}\Gamma\left(10/3\right)\zeta\left(10/3\right)m_D^3}{81\sqrt{3}\pi^{11/3}}\lambda^{7/3}.
$$

With (95) and (96) the relaxation time for thermal conduction can be written as [10]

$$
\tau_{\kappa_{ee}} = 3\left[ \frac{\mathscr{P}}{T^2}\left(1 + \beta\right) \right.
$$
$$
\times \left\{ 2\lambda^2 \zeta\left(3\right) + \frac{(2\pi)^{2/3}}{3}\lambda^{8/3}\zeta\left(\frac{11}{3}\right) \right.
$$
$$
\left.\left. \times \Gamma\left(\frac{14}{3}\right) + \frac{\pi^5}{15}\lambda^3 \right\} \right]^{-1}
$$
$$
\times \left(\left[ \frac{\mu^2 T}{3} + \frac{m_D^2 T}{36}\left(\ln\left(\frac{4}{\pi\lambda}\right) + \gamma_E - \frac{6}{\pi^2}\zeta'\left(2\right) - 3\right) \right.\right.
$$
$$
- 40\frac{2^{2/3}\Gamma\left(8/3\right)\zeta\left(8/3\right)m_D^3}{27\sqrt{3}\pi^{7/3}}\lambda^{5/3}
$$
$$
\left.\left. + 560\frac{2^{1/3}\Gamma\left(10/3\right)\zeta\left(10/3\right)m_D^3}{81\sqrt{3}\pi^{11/3}}\lambda^{7/3} \right]\right)^{-1}. \tag{97}
$$

The thermal relaxation time up to NLO terms contains some anomalous fractional powers originated from the transverse interaction. This in turn changes the temperature dependence of $\tau_{\kappa_{ee}}$ nontrivially. The departure from the Fermi liquid result ($\tau_{\kappa_{ee}} \propto 1/T^2$) is evident from the final expression of $\tau_{\kappa_{ee}}$.

In Figures 8, 9, 10, and 11 we have plotted $\kappa_{ee}$ and $\tau_{\kappa_{ee}}$ with $T$ using (95) and (97). From the plots it can be seen that the inclusion of both the medium modified propagator and $\beta$ decreases the value of $\kappa_{ee}$ and $\tau_{\kappa_{ee}}$. It shows strong deviation from the Fermi liquid in both cases. In Figures 9 and 11 it has been shown that $\beta$ reduces both the thermal conductivity and the thermal relaxation time. This has a serious implication on the total electron conductivity $\kappa_e$. In [50] it has been shown that magnetic interaction decreases $\kappa_{ee}$ which in turn increases the electron-electron collision frequency. Thus to the total electron thermal conductivity electron-electron scattering dominates over electron-ion scattering. The phase space correction due to the medium modified electron dispersion relation further enhances the electron-electron collision frequency [10].

## 7. Conclusion

Investigation of the quasiparticle excitation in ultradegenerate plasma has been the cardinal focus of the present review. The main interesting feature which has been exposed here is the role of magnetic interaction mediated by the transverse gauge bosons leading to the phenomena very different from its high temperature counterpart. In particular the NFL behaviour of various quantities like $C_v$, $\kappa_{ee}$, and $l_{mean}^{total}$ involving excitations near the Fermi surface is a very special characteristic behaviour of degenerate or ultradegenerate plasma. We also show clearly how the dynamical screening leads to finite damping rate emanating from the restricted phase space driven by the Pauli blocking. Besides the damping rate we also present results for the drag and the diffusion coefficients, that is, the energy and the momentum relaxation of the quasiparticle excitation in such a plasma revealing further the domination of the magnetic interaction over the electric interactions in ultradegenerate plasma. This separation between the longitudinal and the transverse sectors is a very special feature of the cold and dense plasma which is not seen in the finite temperature case.

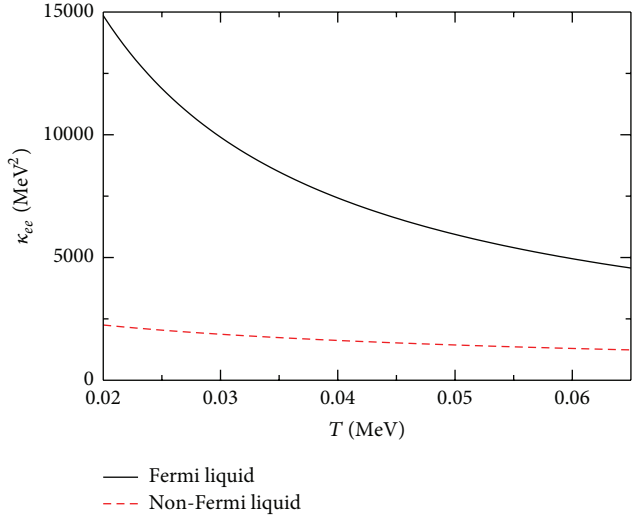

FIGURE 8: Temperature dependence of the electron thermal conductivity and comparison between the Fermi liquid and the non-Fermi-liquid NLO results where $\beta \neq 0$.

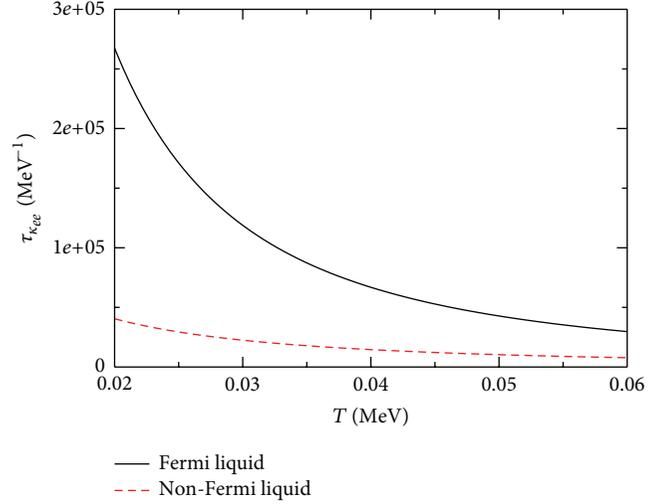

FIGURE 10: Temperature dependence of the thermal relaxation time and comparison between the Fermi liquid and the non-Fermi-liquid NLO results where $\beta \neq 0$.

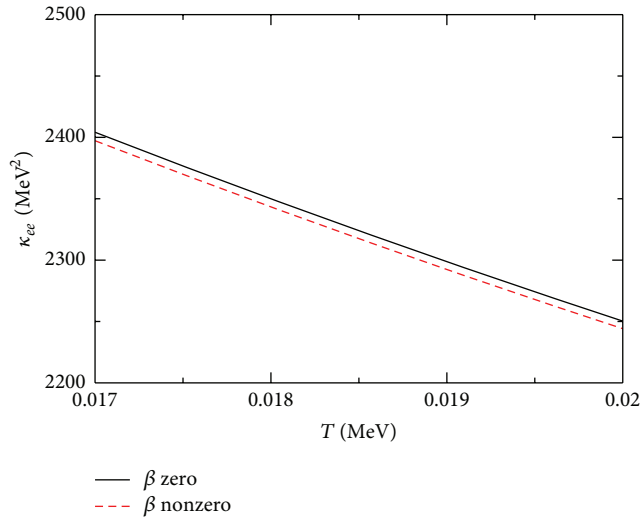

FIGURE 9: The reduction of $\kappa_{ee}$ with the inclusion of $\beta$.

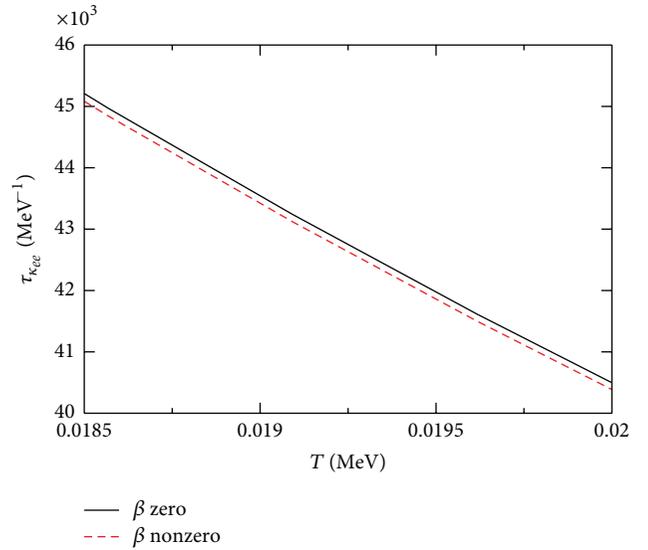

FIGURE 11: The reduction of $\tau_{\kappa_{ee}}$ with the inclusion of $\beta$.

Another aspect of the present work has been the modification of the quasiparticle dispersion relation which changes the phase-space in a nontrivial way leading to modification of $\tau_{\kappa_{ee}}$ and $l_{\text{mean}}$. Further investigations along this line are pertinent to understand various phenomena of astrophysical interests. Some of these extensions have recently been undertaken by various authors showing the importance of this emerging area particularly due to the possibility of experiments in the domain of high chemical potential and low temperature in addition to its relevance in the astrophysical context.

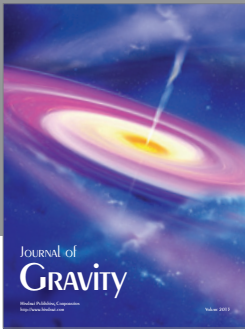

Journal of
**GRAVITY**

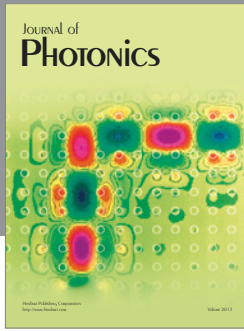

Journal of
**PHOTONICS**

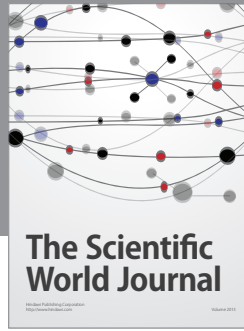

The Scientific
World Journal

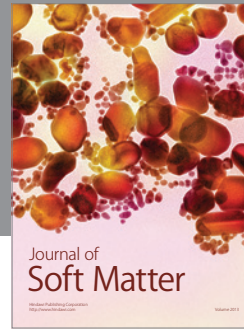

Journal of
**Soft Matter**

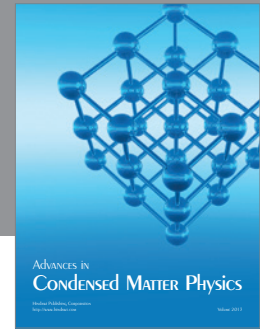

Advances in
Condensed Matter Physics

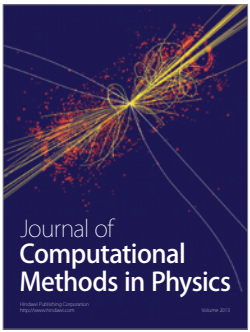

Journal of
**Computational
Methods in Physics**

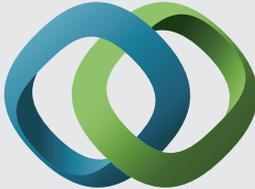

Hindawi

Submit your manuscripts at
http://www.hindawi.com

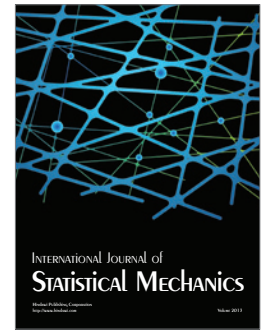

International Journal of
**Statistical Mechanics**

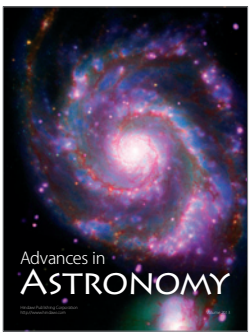

Advances in
**ASTRONOMY**

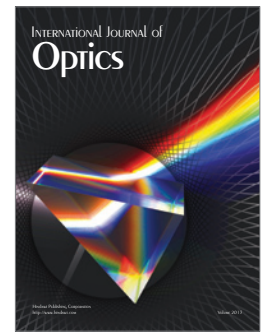

International Journal of
**Optics**

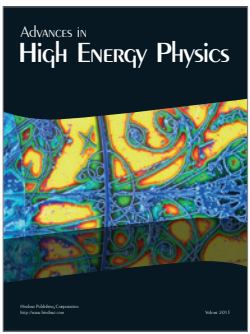

Advances in
High Energy Physics

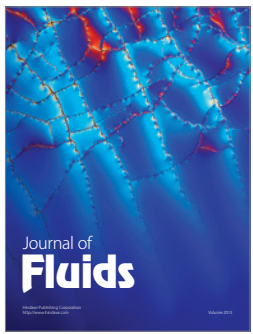

Journal of
**Fluids**

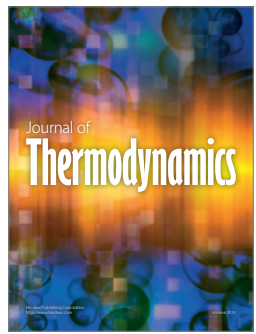

Journal of
**Thermodynamics**

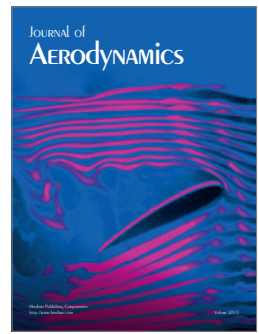

Journal of
**Aerodynamics**

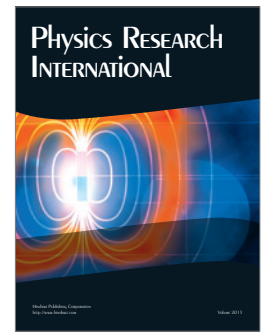

**PHYSICS RESEARCH
INTERNATIONAL**

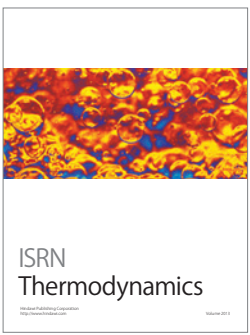

ISRN
Thermodynamics

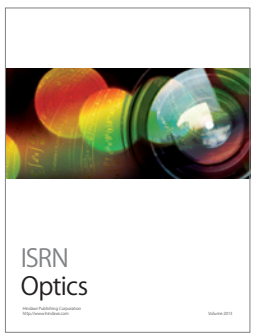

ISRN
Optics

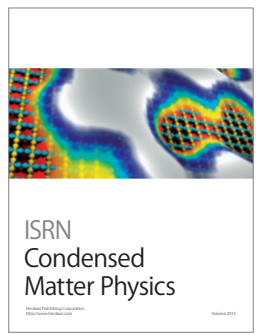

ISRN
Condensed
Matter Physics

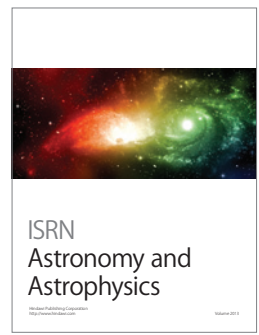

ISRN
Astronomy and
Astrophysics

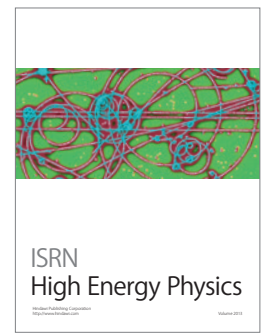

ISRN
High Energy Physics